\documentclass[aps,prb,superscriptaddress,twocolumn]{revtex4-1}
\usepackage{amsmath}
\usepackage{amsfonts,amssymb}
\usepackage{graphicx}
\usepackage{epstopdf}
\usepackage{hyperref}
\usepackage{float}
\usepackage{color}
\usepackage{dsfont}

\begin{document}
	\title{Dephasing-induced Quantum Hall Criticality in the Quantum Anomalous Hall system}
	\author{Fei Yang}
	\affiliation{State Key Laboratory of Low Dimensional Quantum Physics, Department of Physics, Tsinghua University, Beijing 100084, China}
	\author{Dong E. Liu}
	\email{dongeliu@mail.tsinghua.edu.cn}
	\affiliation{State Key Laboratory of Low Dimensional Quantum Physics, Department of Physics, Tsinghua University, Beijing 100084, China}
	\thanks{Corresponding author}
	\begin{abstract}
        Conventional wisdom holds that static disorder is indispensable to the integer quantum Hall effect, underpinning both quantized plateaus and the plateau–plateau transition. We show that pure dephasing, without elastic disorder, is sufficient to generate the same $\theta$-driven criticality. Starting from a Keldysh formulation, we derive an open-system nonlinear $\sigma$ model (NL$\sigma$M) for class A with a topological $\theta$ term but no Cooperon sector, and we demonstrate that nonperturbative instantons still govern a two-parameter flow of $(\sigma_{xx},\sigma_{xy})$. Evaluating $\theta$ in a dephasing quantum–anomalous-Hall setting, we predict a quantum Hall critical point at $\sigma_{xy}=1/2$ with finite $\sigma_{xx}$—the hallmark of the integer quantum Hall universality class realized without Anderson localization. Boundary-driven simulations of the Qi–Wu–Zhang model with local dephasing confirm this prediction and provide an experimentally aligned protocol to extract $(\sigma_{xx},\sigma_{xy})$ from Hall-potential maps. By establishing dephasing as a self-contained route to Hall criticality, our framework reframes plateau physics in open solid-state and cold-atom platforms and offers practical diagnostics for topological transport in nonunitary matter.	
    \end{abstract}
	
	\maketitle
\section{Introduction}
Dephasing, ubiquitous across quantum platforms, governs the quantum-to-classical crossover \cite{quantum_2_classical,quantum_2_classical_1,quantum_2_classical_2} and can generate novel many-body behavior \cite{optical_lattice_rev,dephasing_many_body_1,dephasing_many_body_2}. Fundamentally, it is the weak-measurement–averaged dynamics of a system \cite{continuous_measure}, this placing dephasing at the center of mixed-state topology \cite{mixed_topology,mixed_topology_1,mixed_topology_2,mixed_topology_3,mixed_topology_4} and topological transport in open systems \cite{topology_transport}. Intriguingly, the effect of dephasing is similar to that of disorder, continuous monitoring can drive phase transitions \cite{CM_phase_transition_0,CM_phase_transition,CM_phase_transition_1,CM_phase_transition_2,CM_phase_transition_3} reminiscent of many-body localization \cite{MBL}; the corresponding effective description shares the non-linear $ \sigma $-model (NL$ \sigma $M) structure of Anderson localization \cite{NLSM_CM_1,NLSM_CM_2,NLSM_CM_3}. Moreover, dephasing tight-binding models become diffusive in the thermodynamic limit \cite{dephasing_enhenced}, with a diffuson NL$ \sigma $M as the field theory \cite{NLSM_dephasing}. 

However, disorder and dephasing are fundamentally different: the former is static and unitary, whereas the latter is dynamical and non-unitary. Non-unitarity explicitly breaks time-reversal symmetry and forbids the cooperon, the time-reversed partner of the diffuson, so weak localization is absent in dephasing-dominated dynamics \cite{NLSM_dephasing}. Localization under weak measurement is then attributable to the quantum Zeno effect \cite{quantum_zeno}, not interference physics. These observations suggest that quantum Hall responses in a purely dephasing environment (without disorder) may differ qualitatively from conventional, disordered quantum Hall systems. Since dephasing is now a tunable control parameter in solids and ultracold atoms, resolving the quantum Hall effect under pure dephasing is both timely and essential. Prior studies examined dephasing in disordered quantum Hall settings \cite{dephasing_TIs,dephasing_QHE}, yet the disorder-free, pure-dephasing regime remains unexplored. Intuitively, diffusion in dephasing systems is described by an NL$\sigma$M, while the universal structure of the integer quantum Hall effect is encoded by an NL$\sigma$M with a topological $\theta$-term. Thus, a central open question is whether the $\theta$-term and its nonperturbative phenomena survive in a non-unitary, open setting.
	
In this work, we resolve this issue by investigating a dephasing quantum anomalous Hall (QAH) system, in which an open-system NL$\sigma$M with a topological $\theta$ term is derived. Unlike static disorder, dephasing is inelastic and non-unitary yet leaves the $\theta$ sector intact, so instantons and the Pruisken's two-parameter flow survive in an open setting. Nonperturbative instantons generate a renormalization-group (RG) flow, enabling a $\theta$-controlled quantum phase transition. We identify a critical dephasing rate $\gamma_c$, that separates a robust topological phase ($\gamma<\gamma_c$) from a trivial one ($\gamma>\gamma_c$). The two phases are distinguished by their RG trajectories and finite-size scaling, verified in a boundary-driven Qi--Wu--Zhang (QWZ) model. Numerically, dephasing tight-binding dynamics is diffusive rather than Anderson localized, showing that open-system Hall transport differs qualitatively from disorder-driven transport. We show that dephasing or a weak-measurement/monitoring channel provides a self-contained pathway to integer quantum Hall criticality via the $\theta$-sector.
	
\section{Fermionic gases under dephasing}
Microscopically, dephasing typically arises from inelastic scattering (e.g., electron--phonon interactions) \cite{origin_dephasing,origin_dephasing_1}, which progressively suppresses phase coherence. More broadly, dephasing is an intrinsic feature of open-system dynamics. For fermionic gases it can be captured phenomenologically by B\"{u}ttiker probes \cite{Buttiker_prob} or, more systematically, by the particle-number-conserving Lindblad master equation \cite{dephasing_many_body,dephasing_enhenced,continuous_measure}
\begin{equation}\label{Lindblad_eq}
\frac{d}{dt}\rho(t) =  -i[\hat{H},\rho(t)] - \frac{\gamma}{2}\sum_{j}\left[ \hat{n}_j ,\left[ \hat{n}_j, \rho(t)\right] \right].
\end{equation}
Here $\hat n_j=c_j^\dagger c_j$ denotes the local density and $\gamma$ the dephasing rate. The main structure is the double-commutator (recycling) term on the right-hand side, $\hat n_j \rho(t)\hat n_j$, which, in stochastic unravelings or the Keldysh field theory, generates multiplicative noise and effective quartic couplings. These induce mesoscopic fluctuation effects that, at the level of transport and correlations, closely parallel those produced by static disorder.

Recent work shows that these mesoscopic fluctuations are captured by an NL$\sigma$M \cite{NLSM_dephasing} with action
\begin{equation}
S_0[Q] = \frac{\sigma_{xx}}{8}\!\int d^d x\, \mathrm{Tr}\!\left[\partial_i Q(x)\,\partial_i Q(x)\right],
\end{equation}
where the matrix field $Q(x)$ represents the diffuson Goldstone modes arising from causal-symmetry breaking. This mirrors the disordered NL$\sigma$M, except that dephasing explicitly breaks time-reversal symmetry, eliminating the cooperon (the time-reversed partner of the diffuson). Consequently, weak (anti)localization from quantum interference is absent, and the longitudinal conductivity, $\sigma_{xx}\!\propto\!1/\gamma$, is scale-invariant.

\section{Effective filed theory of QAH under dephasing}
The experimental QAH discovery in magnetically doped topological insulators \cite{observation_QAH} established that quantized Hall physics need not rely on Landau levels but can instead robustly emerge from time-reversal breaking together with strong spin–orbit coupling. Absence of time-reversal symmetry explicitly places the ferimoic systems in class A, whose NL$\sigma$M admits a topological $\theta$-term in even spatial dimensions \cite{anderson_transition}. This term provides the effective field theory of the integer quantum Hall effect and also of disordered QAH systems \cite{disorder_qwz}. 
	
Here, we theoretically examine the QAH system under dephasing, that a most simple QAH system---the QWZ model \cite{qwz_model} is studied, whose effective action is an NL$\sigma$M with $\theta$-term. The QWZ model is defined by the Bloch Hamiltonian $ H(\mathbf{k}) = h_a(\mathbf{k}) \sigma^a $, where $ \vec{h}(\mathbf{k}) = ( \sin k_x, \sin k_y, u +\cos k_x + \cos k_y ) $, the Pauli matrix structure is defined in $ A/B $-sublattices space, $ u $ is the staggering potential. Its topological property is captured by the Chern number $ C = \frac{1}{4\pi}\int_{BZ} dk_x dk_y \vec{n} \cdot \left( \vec{n} \times \vec{n} \right) $, where the unit vector is $ \vec{n} = \vec{h}/|\vec{h}| $, i.e., $ C $ counts the times that Pauli vector $ \vec{h} $ winding around the unit sphere. Nonzero Chern number exists when $ |u|<2 $: To be exact $ C = -1 $ for $ 0<u<2 $ and $ C = 1 $ for $ -2<u<0 $ correspondingly.	 However, the topology of the system under dephasing is no longer characterized by the momentum space invariants, an effective field theory defined in real space is required. 
	
With the Keldysh contour integral of the Lindblad master equation in Eq. (\ref{Lindblad_eq}), the action of fermionic gases under dephasing is obtained as  (see Supplemental Material, Sec. I)
	\begin{equation}\label{Keldysh_action}
		S = \int_{-\infty}^{\infty} dt_1 dt_2  \bar{\Psi }_{t_1} \tilde{G}_0^{-1}(t_1, t_2) \Psi_{t_2} + \frac{i\gamma}{2} \int dt \bar{\Psi }_{t} \Psi_{t}\bar{\Psi }_{t} \Psi_{t},
	\end{equation}
where $ \bar{\Psi} = \left( \bar{\psi}_1, \bar{\psi}_2 \right) $ is defined in the retarded and advanced space. And the nonequilibrium Green's function is
	\begin{eqnarray}
		\tilde{G}_0(\varepsilon,\mathbf{k}) &=& \begin{pmatrix}
			\tilde{G}^{R}(\varepsilon,\mathbf{k})\,\,\, & \tilde{G}^{K}(\varepsilon,\mathbf{k})\\
			0 & \tilde{G}^{A}(\varepsilon,\mathbf{k})
		\end{pmatrix},
	\end{eqnarray}
where $ \tilde{G}^{R/A}(\varepsilon,\mathbf{k}) = \left(\varepsilon - \xi_\mathbf{k} \pm i\frac{\gamma}{2}\right)^{-1} $, $ \xi_\mathbf{k} $ is the spectrum of system. The Keldysh Green's function is determined by the boundary conditions, i.e., $\tilde{G}^{K} = G^R \circ F_\varepsilon  - F_\varepsilon \circ  G^A$, where $ F_\varepsilon = 1 - 2 n_F(\varepsilon) $, $ n_F(\varepsilon) $ is the Fermi-Dirac distribution function. 

The hallmark of the Keldysh action in Eq.~(\ref{Keldysh_action}) is its time-local quartic interaction, which generates interference-driven mesoscopic fluctuations and diffusive dynamics, closely paralleling disordered metals \cite{NLSM_dephasing}. We decouple this term via a Hubbard--Stratonovich field in the density (diffuson) channel and derive the corresponding effective theory for the dephasing QWZ model (technical details presented in the Supplementary Materials, Sec. I). Within a static mean-field treatment the saddle satisfies
$Q^{\alpha\beta}(\mathbf{x};t,t)=Q_0=\tau_3/2$
(with $\tau_3$ acting in Keldysh space). The resulting dephasing-dressed Green's functions are

	\begin{equation}
		G^{R/A}(\varepsilon,\mathbf{k}) = \frac{1}{\varepsilon - \xi_\mathbf{k} \pm i\gamma},
	\end{equation}
so the self-energies are purely imaginary and frequency-independent, \( \Sigma^{R/A}=\mp i\gamma \). This is a natural consequences of time-local interaction, that dephasing only broadening the energy bands, while leaves its energy level invariant. Thus, topological Anderson insulator which results from the band-inversion due to the disorder \cite{TAI} is not expected in the dephasing topological insulators. Nevertheless, the broadening of energy bands would leads to trivialization of band topology,  while a theoretical framework of this process remains unknown.
	
In response to this challenge, we introduce quantum fluctuation to the system, $ Q(x) = T(x) Q_0 T^{-1}(x) \in U(2)/\left[U(1)\times U(1) \right]$, then gradient expansion leads to the effective action of these soft modes in the dephasing QWZ model, which is a NL$\sigma$M with $ \theta $-term (or the Pruisken's action) \cite{Pruisken_action}
	\begin{equation}\label{pruisken_action}
		S_{\text{eff}}[Q] = S_0[Q] + \frac{\theta}{16\pi} \int d^2x \text{Tr}\left( \epsilon_{ij} Q \partial_i Q \partial_j Q \right), 
	\end{equation}
where the topological angle $ \theta = \theta_1 + \theta_2 $, and $ \theta_1 $ and $ \theta_2 $ are defined as follows  (see Supplemental Material, Sec. II)
	\begin{eqnarray} 
		\label{theta1}
		\theta_1 &=& 8\gamma   \int \frac{d^2k}{4\pi^2} D_E^+ D_E^- F_k,\\
		\label{theta2}
		\theta_2 &=& 4\pi i \int_{-\infty}^{E} d\varepsilon \int \frac{d^2k}{4\pi^2}\left( D_\varepsilon^{+2} - D_\varepsilon^{-2}\right) F_k,
	\end{eqnarray}
that $ F_k = \epsilon_{abc} \partial_1 h_a \partial_2 h_b h_c $ is related to the Berry curvature of QWZ model, and $ D^{\pm}_\varepsilon = \frac{1}{\left( \varepsilon \pm i\gamma \right)^2 - h_a h^a} $ is related to its Green's function. Generally speaking, $ \theta_1 $ is responsible to the Hall conductance at the energy level $ E $,  and $ \theta_2 $ is responsible to the Hall conductance of all energy levels below $ E $. We find that the topological angle $\theta$ is parameterized by $(E,u,\gamma)$, i.e., $\theta(E,u,\gamma)$. And the Hall conductivity $\sigma_{xy}=\theta/2\pi$ is a topological invariant (or the Chern number): the winding number of the map with which the real-space two-dimensional torus $T^2$ covers the soft-mode target manifold $U(2)/\left[U(1)\times U(1) \right]$.
	
The second term in Pruisken’s action is the $\theta$-term $ S_{\theta} = \frac{\theta}{16\pi}\int d^2x \text{Tr}\left( \epsilon_{ij} Q \partial_i Q \partial_j Q \right) =  i\theta  $, which contributes a phase factor $e^{i\theta}$. The periodicity
$e^{i\theta}=e^{i(\theta+2\pi n)}$ implies $n$ topologically inequivalent yet dynamically equivalent ground states. This topological degeneracy produces non-perturbative tunneling captured by instantons~\cite{instanton_ref,instanton_ref_1}. Within the dilute-instanton gas approximation, the two-parameter $\beta$-functions for $(\sigma_{xx},\sigma_{xy})$ reproduce the universal features of the integer quantum Hall effect~\cite{RG_QHE,RG_QHE_1}. However, the term $- \frac{1}{2\pi^2 \sigma_{xx}} $ in $ \beta_{xx} $ reflecting the weak localization effect should be discarded for the purely dephasing quantum Hall systems, then the corresponding $ \beta $-functions become
	\begin{eqnarray}
		\label{beta_xx}
		\beta_{xx} &=& \frac{\partial \sigma_{xx}}{\partial \ln L}= -D_0\sigma_{xx}^2 e^{-2\pi \sigma_{xx}}\cos(2\pi\sigma_{xy}),\\
		\label{beta_xy}
		\beta_{xy} &=& \frac{\partial \sigma_{xy}}{\partial \ln L}= - D_0\sigma_{xx}^2e^{-2\pi \sigma_{xx}}\sin(2\pi\sigma_{xy}),
	\end{eqnarray}	
where $ D_0 >0 $ is a numerical constant.

At the Hall plateaus $\sigma_{xy}=n$ ($n\in\mathbb{Z}$), one has $\beta_{xy}=0$, so $\sigma_{xy}$ is scale-invariant. In our flow equations, $\beta_{xx}<0$ at $\sigma_{xy}=n$, which tends to decrease $\sigma_{xx}$ under coarse-graining. However, the dilute-instanton-gas approximation is controlled only for moderately large $\sigma_{xx}$ (small instanton fugacity $e^{-2\pi\sigma_{xx}}$) and breaks down as $\sigma_{xx}\to 0$~\cite{RG_QHE}. In addition, a purely dephasing QAH system lacks the coherent-interference mechanisms required for Anderson localization, and numerical studies indicate diffusive transport in the thermodynamic limit~\cite{dephasing_enhenced}. It is therefore natural to expect $\sigma_{xx}$ to saturate at a finite value, yielding a plateau fixed point $(\sigma_{xy},\sigma_{xx})=(n,\sigma_0)$ with $\sigma_0>0$.

\begin{figure}
		\includegraphics[width = 0.5\textwidth]{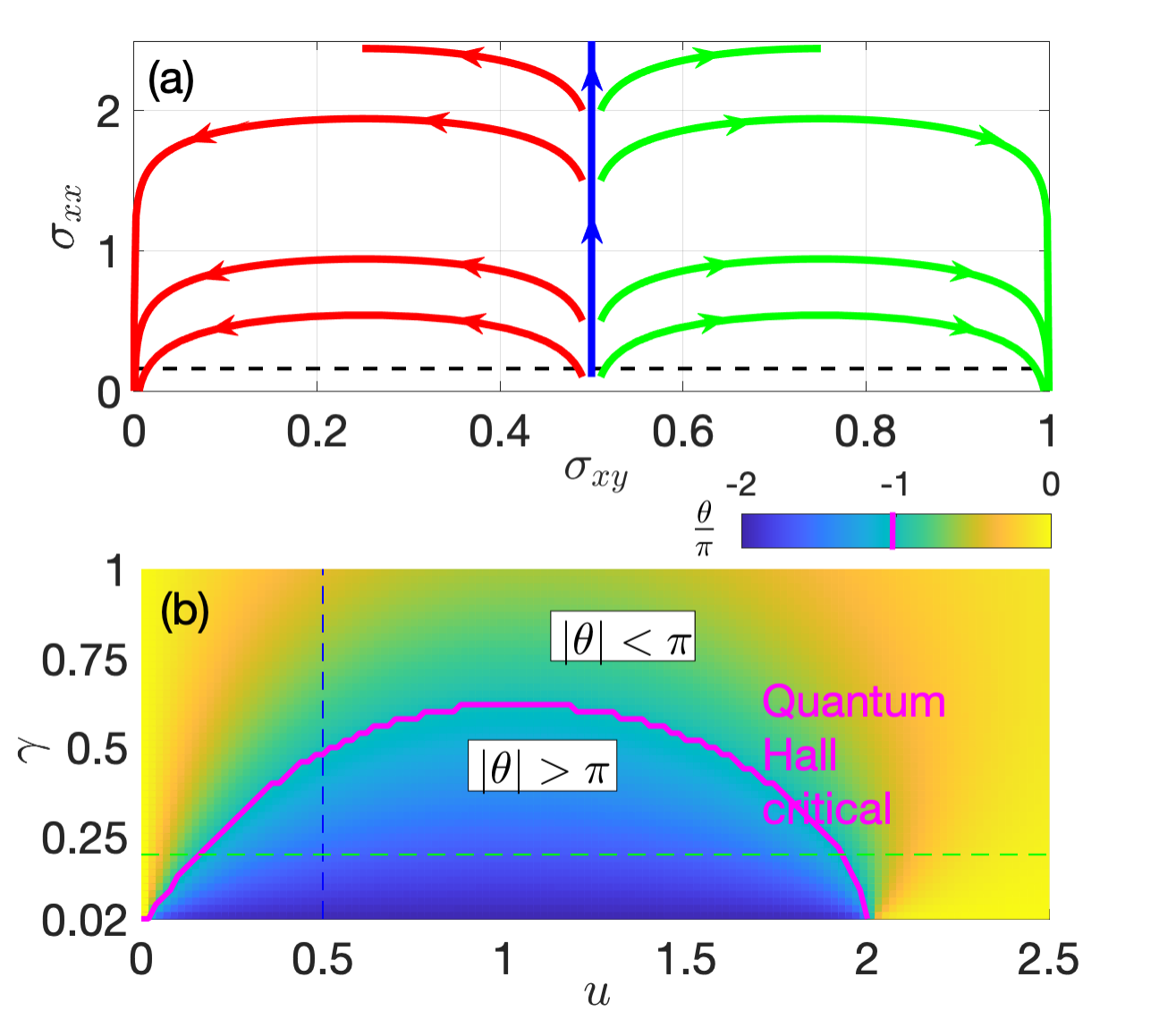}
		\caption{(a) The trajectories of two parameter renormalization group flow of dephasing quantum anomalous hall systems, where the numerical constant $D_0=2$. (b) The phase diagram of dephasing Qi-Wu-Zhang model for $ E = 0 $, in which there are two phases, denoted as $ |\theta|>\pi $ and $ |\theta|<\pi $ correspondingly, where $ \gamma $ is the rate of dephasing, $ u $ is the staggering potential of the model, $ \theta = \theta_1 + \theta_2 $ is the topological angle. We observed a dephasing-indeced quantum phase transition when the band topology is nontrivial (that $ |u|<2 $), where the quantum Hall critical points are those $ |\theta|=\pi $ (mengta-solid line).}
		\label{phase_diag_QWZ}
	\end{figure}

At the quantum Hall critical points $(\sigma_{xy},\sigma_{xx})=(n+\tfrac{1}{2},\sigma_{xx}^*)$, $\beta_{xy}=0$ pins $\sigma_{xy}=n+\tfrac{1}{2}$ at all scales. Formally, our $\beta_{xx}>0$ there, which naively drives $\sigma_{xx}\to\infty$; yet the prefactor $\sigma_{xx}^2 e^{-2\pi\sigma_{xx}}$ suppresses the flow at large $\sigma_{xx}$, so $\beta_{xx}\to 0$ and $\sigma_{xx}$ approaches a finite constant $\sigma_{xx}^*$. Thus the transition occurs along a critical line with $\sigma_{xy}=n+\tfrac{1}{2}$ and finite $\sigma_{xx}=\sigma_{xx}^*$.
Away from criticality, if $n+\tfrac{1}{2}<\sigma_{xy}<n+1$ then $\beta_{xy}>0$ and $\sigma_{xy}\to n+1$; if $n<\sigma_{xy}<n+\tfrac{1}{2}$ then $\beta_{xy}<0$ and $\sigma_{xy}\to n$. In both regimes, $\beta_{xx}<0$ at large scales and $\sigma_{xx}$ flows toward $\sigma_0$. These RG trajectories are illustrated in Fig.~\ref{phase_diag_QWZ}(a). Physically, a finite longitudinal conductivity is expected in dephasing-dominated QAH systems, while $\sigma_{xy}$ remains quantized on the plateaus, and $\sigma_{xx}$ attains its maximal value near the quantum Hall critical line $\sigma_{xy}=n+\tfrac{1}{2}$.

The phase diagram of the dephasing QWZ model is shown in Fig.~\ref{phase_diag_QWZ}(b). When the band topology is nontrivial ($|u|<2$), the topological angle $\theta$ evolves from $|\theta|>\pi$ to $|\theta|<\pi$ as the dephasing rate $\gamma$ increases. The dephasing-induced quantum Hall critical line is defined by $|\theta(0,u,\gamma)|=\pi$ (magenta solid line in Fig.~\ref{phase_diag_QWZ}(b)). For fixed $u$, there is a single critical dephasing rate $\gamma_c$ satisfying $|\theta(0,u,\gamma_c)|=\pi$ (blue dashed line). For fixed $\gamma$, there are two critical values of the staggering potential, $u_1$ and $u_2$, such that $|\theta(0,u_{1},\gamma)|=|\theta(0,u_{2},\gamma)|=\pi$ (blue dashed line). At large scales, the RG flow sends the topologically nontrivial regime ($|\theta|>\pi$) to the fixed point $|\theta|=2\pi$ and the trivial regime ($|\theta|<\pi$) to the fixed point $|\theta|=0$. This prediction can be tested via finite-size scaling.
	
\section{Numerical studies of boundary driven QWZ model under dephasing}
\begin{figure}
		\includegraphics[width = 0.5\textwidth]{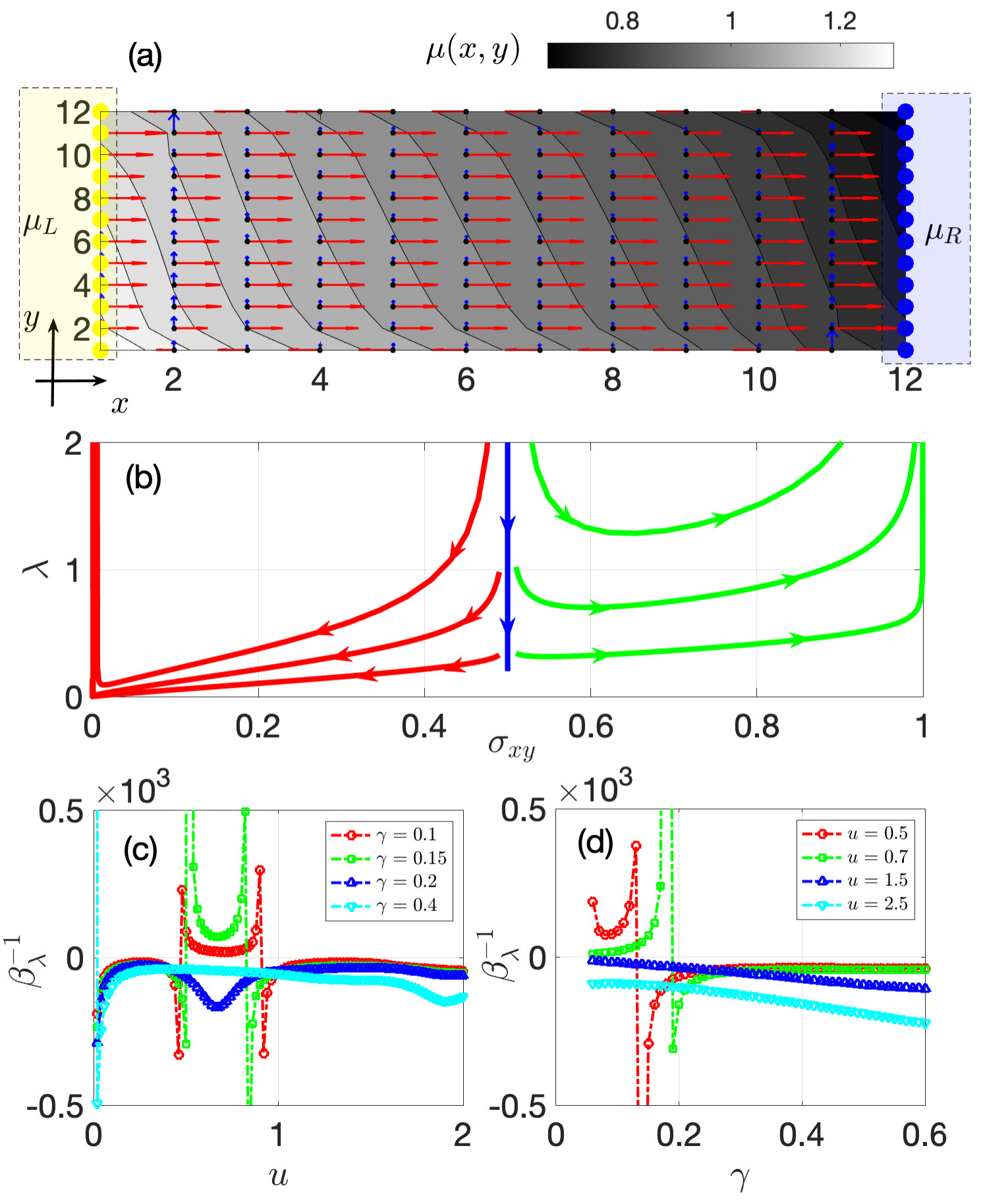}
		\caption{(a) The contour plot of chemical potential $\mu(x,y)$ and the current distribution in the steady-state of the boundary driven Qi-Wu-Zhang model with dephasing, that the system is in open boundary conditions and the source is marked in yellow while drain is marked in blue, where $ \gamma= 0.2$, $ u=0.6 $ and $ L=12 $. The coupling between the system and source/drain is set to a constant $\Gamma=0.1$. (b) The trajectories of two parameters $(\sigma_{xy},\lambda)$ renormalization group flow, it obvious that $ \beta_\lambda<0 $ for $ \sigma_{xy}\leq1/2 $, while $ \beta_\lambda>0 $ can only be satisfied when $ \sigma_{xy}> 1/2 $, where the numerical constant $ D_0=2 $. (c) The inverse $ \beta $-function $ \beta_\lambda^{-1} $ versus $u$ for $ \gamma=0.1, 0.15, 0.2, 0.4 $, (d) The inverse $ \beta $-function $ \beta_\lambda^{-1} $ versus $ \gamma $ for $ u=0.5, 0.7, 1.5, 2.1 $, where $ \beta_\lambda $ is the fitting parameter of the function $ \lambda(L) = \beta_\lambda \ln L + \alpha$ for $ L = 8, 12, 16, 20 $.}
		\label{chemical_RG_num}
\end{figure}

We numerically compute the nonequilibrium steady-state of the boundary-driven QWZ model on a square lattice with $L_x=L_y=L$. The left boundary ($x=1$) is coupled to a source reservoir at chemical potential $\mu_L$, and the right boundary ($x=L$) is coupled to a drain reservoir at $\mu_R$; the system–reservoir coupling is spatially uniform with strength $\Gamma=0.1$ [Fig.~\ref{chemical_RG_num}(a)]. Dephasing is modeled by a particle-number-conserving Lindblad master equation with a site-independent rate $\gamma$ acting identically on both sublattices $A$ and $B$ (see Supplemental Material, Sec. IV).

In the boundary-driven Hall setup, the electrochemical potential in the bulk can be modeled as an electrostatic field $\mu(x,y)$ that satisfies Laplace’s equation, $\nabla^{2}\mu(x,y)=0$, away from the contacts. Imposing no-flux through the transverse edges $y=1$ and $y=L$ (e.g. $J_y(x,1)=J_y(x,L)=0$) and using the Hall conductivity tensor yields the condition \cite{boundary_eq_Hall,source_drain_anomalous_Chern}:
\begin{equation}\label{chemical_func}
		\frac{\partial}{\partial y} \mu(x,y)  = \lambda \frac{\partial }{\partial x} \mu(x,y),
\end{equation}
where $\lambda=\sigma_{xy}/\sigma_{xx}$. Numerical values of $ \mu(x,y) $ in boundary driven QWZ model is presented in Fig. \ref{chemical_RG_num} (a), the computed current densities $J_x(x,y)$ and $J_y(x,y)$ show small deviations from the ideal boundary condition $J_y(x,1)=J_y(x,L)=0$, attributable to the chiral edge modes which localized at the boundary of system and the finite-size effect.

The dephasing QWZ model exhibits a nontrivial two-parameter renormalization  flow in $(\sigma_{xx},\sigma_{xy})$. In particular, $\sigma_{xy}$ shows distinct finite-size scaling on the intervals $n<\sigma_{xy}<n+\tfrac{1}{2}$ and $n+\tfrac{1}{2}<\sigma_{xy}<n+1$ ($n\in\mathbb{Z}$), whereas $\sigma_{xx}$ scales similarly across both. Consequently, the Hall-angle ratio $\lambda=\sigma_{xy}/\sigma_{xx}$ displays piecewise scaling, as illustrated by the RG trajectories in the $(\sigma_{xy},\lambda)$ plane [Fig.~\ref{chemical_RG_num}(b)]. Numerically, we find $\beta_\lambda\equiv d\lambda/d\ln L>0$ only for $\sigma_{xy}>1/2$, while $\beta_\lambda<0$ for $\sigma_{xy}\le 1/2$. This sign structure enables an efficient finite-size-scaling verification of the phase diagram in Fig.~\ref{phase_diag_QWZ} using $\lambda$ alone. Operationally, $\lambda$ is directly obtained from the electrochemical-potential profile: computing coarse-grained gradients $\mu_x\equiv \partial_x\mu$ and $\mu_y\equiv \partial_y\mu$ yields $\lambda=\mu_y/\mu_x$, (see Supplemental Material, Sec. V).

Numerical value of inverse $ \beta $-function $ \beta_\lambda^{-1} $ is presented in Fig. \ref{chemical_RG_num} (c, d), where $ \beta_\lambda^{-1} $ is obtained by fitting the function $ \lambda(L) = \beta_{\lambda} \ln L + \alpha$ for various values of system size $L$. In Fig. \ref{chemical_RG_num} (c), we analyze $ \beta_{\lambda}^{-1} $ versus $ u $. For $ \gamma=0.10 $ and $ 0.15 $, we find two critical values of $ u $ , denoted as $ u_1 $ and $ u_2 $ ($u_1<u_2$), that $ \beta_{\lambda}^{-1} >0 $ in between these two points ( i.e., $ u_1<u<u_2 $), while $ \beta_{\lambda}^{-1} <0 $ otherwise. Theoretically, this means that the steady-state is in topological non-trivial phases ($ |\theta|>\pi $) when $ u_1<u<u_2 $, otherwise, it is in topological trivial ($ |\theta|>\pi $). For larger dephasing, e.g., $\gamma=0.20$ and $0.40$, we find $\beta_\lambda^{-1}<0$ for all $u$, consistent with a trivial phase ($|\theta|<\pi$). These trends quantitatively agree with the green dashed boundary in Fig.\ref{phase_diag_QWZ}(b).

In Fig.~\ref{chemical_RG_num}(d), we examine $\beta_\lambda^{-1}$ as a function of the dephasing rate $\gamma$. For $u=0.5$ and $u=0.7$, a critical value $\gamma_c$ emerges: $\beta_\lambda^{-1}>0$ (topologically nontrivial, $|\theta|>\pi$) for $\gamma<\gamma_c$, while $\beta_\lambda^{-1}<0$ (topologically trivial, $|\theta|<\pi$) for $\gamma>\gamma_c$. By contrast, for $u=1.5$ and $u=2.7$ we find $\beta_\lambda^{-1}<0$ for all $\gamma$, indicating a trivial phase throughout. These results quantitatively agree with the blue dashed boundary in Fig.~\ref{phase_diag_QWZ}(b). 

Residual discrepancies in Fig.~2 likely stem from the boundary-driven protocol probing transport at a reference energy $E\neq 0$ (often inside a valence or conduction band), which reduces the topological angle $\theta(E)$ relative to its midgap value at $E=0$, thereby compressing the nontrivial region and producing a modest mismatch between continuum expectations and lattice numerics  (see Supplemental Material, Sec. III).
	
\section{Discussion}
We develop the effective field theory of a dephasing-driven QAH system: a class-A NL$\sigma$M that retains a topological $\theta$-term. While dephasing is often said to ``act like'' disorder because both generate diffusion, the two cases differ qualitatively: static disorder enables quantum-interference corrections (weak-(anti)localization), whereas pure dephasing suppresses phase coherence, removing the weak-(anti)localization channel. Consequently, the longitudinal conductivity remains finite and dephasing-set (microscopically $\sigma_{xx}\propto 1/\gamma$) rather than flowing to zero under RG. At the same time, topology survives decoherence: the $\theta$-term supports non-perturbative (instanton) effects and drives a nontrivial two-parameter flow in $(\sigma_{xx},\sigma_{xy})$. Evaluating $\theta$ in the dephasing QWZ model identifies a dephasing-induced critical line given by $|\theta(0,u,\gamma)|=\pi$, along which the Hall conductivity alone changes plateau value at $\sigma_{xy}=n+\tfrac{1}{2}$ while $\sigma_{xx}$ remains finite. Thus, unlike the standard disorder-driven integer-QH transition (localization--delocalization), the transition here is a quantum topological phase transition triggered by dephasing, with quantized $\sigma_{xy}$ on plateaus and scale-insensitive $\sigma_{xx}$ set by $\gamma$.

\section{Experimental suggestion}
Ultracold fermionic gases in optical lattices~\cite{optical_lattice_1,optical_lattice_2} provide a controlled platform to study many-body topology. In these systems, synthetic spin–orbit coupling is engineered via Raman-coupled lattices~\cite{optical_ramann,optical_ramann_2,optical_ramann_3}, and QAH band topology can be diagnosed from quench dynamics~\cite{quench_dynamics,quench_dynamics_1}. In the $^{87}\mathrm{Sr}$ realization of the QWZ model~\cite{experimental_qwz}, the pseudospins are chosen as, e.g., $|\uparrow\rangle=|F=\tfrac{9}{2},m_F=\tfrac{9}{2}\rangle$ and $|\downarrow\rangle=|F=\tfrac{9}{2},m_F=\tfrac{7}{2}\rangle$, which are coherently coupled to emulate spin--orbit physics. To examine QWZ dynamics under controlled dephasing, one can explicitly incorporating weak optical-pumping (``recycling'') pathways from these ground-state sublevels via allowed excited-state hyperfine manifolds $|F',m_F'\rangle$ (set by polarization selection rules). These channels yield number-conserving incoherent scattering and phase noise, potentially spin relaxation depending on polarization/detuning, captured by a Lindblad master equation and tunable via the intensity, detuning, and polarization of a far-off-resonant auxiliary beam~\cite{dephasing_many_body}.

Operationally, (i) open and tune the optical-pumping channels to set a dephasing rate $\gamma$; (ii) calibrate $\gamma$ using Ramsey/echo of the pseudospin and momentum-resolved coherence measurements; and (iii) repeat quench-based topological probes (e.g., dynamical winding of spin textures, mean chiral displacement, or related quench tomographies) while varying $\gamma$. This protocol maps the dephasing dependence of QWZ indicators and tests predictions for topology in open quantum systems, while monitoring atom loss and heating to ensure that dephasing, rather than depletion, is the dominant effect.

\begin{acknowledgements}
\textbf{\emph{Acknowledgements.}}
The authors thank Qinghong Yang, Yu-Dong Wei and Yuanchen Zhao for helpful discussions. This work is supported by the National Natural Science Foundation of China (Grant No.~92365111), Shanghai Municipal Science and Technology (Grant No.~25LZ2600200), Beijing Natural Science Foundation (Grants No.~Z220002), and the Innovation Program for Quantum Science and Technology (Grant No.~2021ZD0302400).
\end{acknowledgements}

\appendix
\section{Keldysh action of fermionic gases under decoherence}
Principle of Keldysh contour path integral is to including the forward/backward branch of time evolution (the dynamics corresponds to "ket" and those of "bra"), that the partition function of Fermionic gases under dephasing is
\begin{eqnarray}
	\label{partition_function}
	Z &=& \text{Tr}\left[ e^{t\mathcal{L}} \rho_0 \right]
	= \int \mathcal{D}\left[\bar{\psi}_{+}, \psi_{+}, \bar{\psi}_{-}, \psi_{-} \right] e^{iS},\\
	S &=&  \int dx \int_{-\infty}^{\infty} dt \left[  \bar{\psi }_+(t) \left( i\partial_t  - H_+ +\frac{i\gamma}{2}\right)\psi_+(t) \right.\nonumber\\
	&&\left.- \bar{\psi }_-(t) \left( i\partial_t  - H_- - \frac{i\gamma}{2}\right)\psi_-(t)\right. \nonumber\\
	&&\left.- i\gamma \bar{\psi }_+(t) \psi_+ \bar{\psi }_-(t) \psi_- (t)\right],
\end{eqnarray}
where the subscript $ +/- $ corresponds to forward/backward branch along the contour.
With a canonical Keldysh-Lakin-Ovchinnikov (KLO) transformation \cite{Keldysh_book,Keldysh_paper} that 
\begin{eqnarray}
	\psi_{+} = \frac{1}{\sqrt{2}}\left( \psi_{1} + \psi_{2} \right),\qquad \psi_{-} = \frac{1}{\sqrt{2}}\left( \psi_{1} - \psi_{2} \right),\\
	\bar{\psi}_{+} = \frac{1}{\sqrt{2}}\left( \bar{\psi}_{1} + \bar{\psi}_{2} \right),\qquad \bar{\psi}_{-} = \frac{1}{\sqrt{2}}\left( \bar{\psi}_{2} - \bar{\psi}_{1} \right),
\end{eqnarray}
then, the action becomes
\begin{eqnarray}
	S &=&  \int dx \int_{-\infty}^{+\infty} dt_1 dt_2 \begin{pmatrix}
		\bar{\psi}_1 & \bar{\psi}_2
	\end{pmatrix}_{t_1} \tilde{G}^{-1}(t_1,t_2)  \begin{pmatrix}
		\psi_{1}\\
		\psi_{2}
	\end{pmatrix}_{t_2} \nonumber\\
	&&+ i \frac{\gamma}{2} \int dt \left[\bar{\psi}_1(t)\psi_{1}(t) + \bar{\psi}_2(t) \psi_{2}(t) \right]^2,\\
	&=& \int_{-\infty}^{\infty} dt_1 dt_2  \bar{\Psi }_{t_1} \tilde{G}_0^{-1}(t_1, t_2) \Psi_{t_2} + i\frac{\gamma}{2} \int dt \bar{\Psi }_{t} \Psi_{t}\bar{\Psi }_{t} \Psi_{t},\nonumber\\
	&=& \tilde{S}_0 + S_{int}
\end{eqnarray}
where $ \bar{\Psi} = \left( \bar{\psi}_1, \bar{\psi}_2 \right) $ is defined in the retarded and advanced space, and the non-equilibrium Green's function is
\begin{eqnarray}
	\tilde{G}(t_1,t_2) &=& \begin{pmatrix}
		\tilde{G}^{R}(t_1,t_2) & \tilde{G}^{K}(t_1,t_2)\\
		0 & \tilde{G}^{A}(t_1,t_2)
	\end{pmatrix} - \frac{i}{2}\tau_1 \delta_{t_1,t_2},\\
	\tilde{G}(\varepsilon,\mathbf{k}) &=&  \begin{pmatrix}
		\tilde{G}^{R}(\varepsilon,\mathbf{k}) & \tilde{G}^{K}(\varepsilon,\mathbf{k})\\
		0 & \tilde{G}^{A}(\varepsilon,\mathbf{k})
	\end{pmatrix} - \frac{i}{2}\tau_1.
\end{eqnarray}
Generally, time-local term $ \frac{i}{2}\tau_1 \delta_{t_1,t_2} $ is discarded because it is a manifold of measure zero and omitting it is inconsequential for most purposes \cite{Keldysh_book}. The Keldysh Green's function is determined through the boundary conditions, that $\tilde{G}^{K} = G^R \circ F_\varepsilon  - F_\varepsilon \circ  G^A$, where $ F_\varepsilon = 1 - 2 n_F(\varepsilon) $, $ n_F(\varepsilon) $ is the Fermi-Dirac distribution function.

Our Keldysh action is different from the rigorous continuum time limit one\cite{NLSM_dephasing}. In their work, the saddle-point equation is identical to the disorder system, and a non-linear $ \sigma $ model (NL$\sigma$M) is derived approximately near the Fermi energy. However, we encounter with a time-local interaction term, $ \bar{\Psi }_{t} \Psi_{t}\bar{\Psi }_{t} \Psi_{t} $, which is different from the disorder system. Then, its necessary to derive the exact saddle-point equation and propose an NL$\sigma$M which independent of Fermi energy. In addressing this issue, we introduce a Hubbard--Stratonovich matrix-field $ Q^{\alpha\beta}(t,t) = \{\Psi_\alpha(t)\bar{\Psi}_\beta(t)\}$, which is time-local and Hermitian, i.e., $ Q^{\alpha\beta *}(t,t) = Q^{\beta\alpha} (t,t)$, then
\begin{eqnarray}
	\mathds{1} &=&  \int\mathcal{D}[Q] \exp\left( -\frac{\gamma}{2}\sum_{\alpha\beta}\left|Q^{\alpha\beta} \right|^2\right) \nonumber\\
	&=& \int\mathcal{D}[Q] \exp\left[-\frac{\gamma}{2} \text{Tr}\left( Q^2\right) \right].
\end{eqnarray}
Performing a shift of matrix field, $ Q^{\alpha\beta} \rightarrow Q^{\alpha\beta} -\Psi_{\alpha}\bar{\Psi}_{\beta}$, then

\begin{eqnarray}
	\mathds{1} &=& \int\mathcal{D}[Q] \exp\left[-\frac{\gamma}{2} \sum_{\alpha\beta}\left(Q^{\alpha\beta}-\Psi_{\alpha}\bar{\Psi}_{\beta}\right)\left(Q^{\beta\alpha}-\Psi_{\beta}\bar{\Psi}_{\alpha}\right)\right],\nonumber\\
	&=& \int\mathcal{D}[Q] \exp\left[-\frac{\gamma}{2} \text{Tr}\left( Q^2 \right) - \gamma\bar{\Psi} Q \Psi + \frac{\gamma}{2} \bar{\Psi}\Psi\bar{\Psi}\Psi \right].
\end{eqnarray}

This directly leads to
\begin{equation}\label{Hubbard-Stronovich}
	\exp\left[-\frac{\gamma}{2}\left( \bar{\Psi} \Psi \right)^2\right] = \int\mathcal{D}[Q]\exp\left[ -\frac{\gamma}{2} \text{Tr}\left(Q^2\right) - \gamma\bar{\Psi } Q \Psi \right],
\end{equation}
in analogous with the molecular field in the ferromagnetism, the time-local interaction term in the action is equivalent to the fermions coupled to a Bosonic matrix field $Q$.
Substitute it into the partition function in Eq. (\ref{partition_function}), we have
\begin{widetext}
\begin{eqnarray}
	Z = \int \mathcal{D}[\psi,\bar{\psi},Q] \exp\left\lbrace  -\int_{-\infty}^{+\infty} dt_1 dt_2 \bar{\Psi }_{t_1} \left[ -i\tilde{G}_0^{-1}(t_1,t_2) + \gamma Q(t_1)\delta_{t_1,t_2} \right] \Psi_{t_2} - \frac{\gamma}{2}\int dt \text{Tr}\left(Q^2(t)\right) \right\rbrace.
\end{eqnarray}
Gaussian integration of Grassmann variables leads to an effective bosonic theory
\begin{eqnarray}
	Z = \int \mathcal{D}[Q] \exp\left[ - \frac{\gamma}{2}\text{Tr}\left(Q^2(t)\right) \right] \det\left[-i\tilde{G}_0^{-1}(t_1,t_2) + \gamma Q(t_1)\delta_{t_1,t_2} \right].
\end{eqnarray}
Then, the action of $ Q $ is
\begin{equation}
	S[Q(t)] = -i\text{Tr}\ln\left[-i\tilde{G}_0^{-1}(t_1,t_2) + \gamma Q(t_1)\delta_{t_1,t_2} \right] +\frac{i\gamma}{2}\text{Tr}\left(Q^2(t)\right).
\end{equation}
\end{widetext}

Taking the variation over $ Q(t) $, one obtains the saddle-point equation as
\begin{equation}
	Q(t_3) =  \text{Tr}\left[ \frac{i\delta_{t_1,t_2}\delta_{t_1,t_3}}{\tilde{G}_0^{-1}(t_1,t_2) + i\gamma Q(t_1)\delta_{t_1,t_2}} \right].
\end{equation}
There are two time argument in $ \tilde{G}_0 (t_1,t_2) $, while only one time argument in $ Q(t_1) $, that we can't diagonalize them simultaneously.
Fortunately, steady-state implies that it is reasonable to assume that $ Q(t)=Q_0 $ is a constant, this gives the \textit{static mean-field approximation}. Then, the saddle-point equation becomes
\begin{equation}
	\label{saddle_eq_in_k}
	Q_0 = \int\frac{d\varepsilon}{2\pi} \sum_{\mathbf{k}} \frac{i}{\tilde{G}_0^{-1}(\varepsilon, \mathbf{k}) + i\gamma Q_0}.
\end{equation}
Generally, the mean-field $ Q_0 $ (or the self-energy) has the structure
\begin{equation}
	Q_0  = \begin{pmatrix}
		q^R & q^K\\
		0 & q^A
	\end{pmatrix},\qquad q^K_\varepsilon = q^R_\varepsilon \cdot F_\varepsilon - F_\varepsilon \cdot q^A_\varepsilon.
\end{equation}
substituting it into Eq. (\ref{saddle_eq_in_k}) leads to
\begin{equation}
	q^R = \int\frac{d\varepsilon}{2\pi} \sum_{\mathbf{k}} \frac{i}{\varepsilon  - \xi_{\mathbf{k}} + \frac{i\gamma}{2}+ i\gamma q^R}.
\end{equation}

\begin{widetext}
Rewritten $ q^R $ as $ q_r + iq_i $, where $ q_r, q_i\in \mathbb{R} $.The integral kernel becomes
\begin{eqnarray}
	I(\varepsilon) &=& \sum_{\mathbf{k}} \frac{i}{\varepsilon - \xi_{\mathbf{k}} + \frac{i\gamma}{2} + i\gamma q^R},\nonumber\\
	&=& \int_{-\infty}^{\infty} d\xi g(\xi) \left[\frac{\gamma \left( \frac{1}{2} + q_r \right)}{\left( \varepsilon - \xi - \gamma q_i \right)^2 + \gamma^2\left( \frac{1}{2} + q_r \right)^2  }+i\frac{\varepsilon - \xi - \gamma q_i}{\left( \varepsilon - \xi - \gamma q_i \right)^2 + \gamma^2\left( \frac{1}{2} + q_r \right)^2  } \right],
\end{eqnarray}
where $g(\xi) $ is the density of states at $ \xi $, which is usually an even function.
It is easy to find that the imaginary part of this integral kernel $ I(\varepsilon) $ satisfy
\begin{eqnarray}
	I_i(\varepsilon) &=& \int_{-\infty}^{\infty} d\xi g(\xi)\frac{\varepsilon - \xi - \gamma q_i}{\left( \varepsilon - \xi - \gamma q_i \right)^2 + \gamma^2\left( \frac{1}{2} + q_r \right)^2  },\\
	&=& \int_{0}^{\infty} d\xi g(\xi)\left[ \frac{\varepsilon - \xi - \gamma q_i}{\left( \varepsilon - \xi - \gamma q_i \right)^2 + \gamma^2\left( \frac{1}{2} + q_r \right)^2  } +  \frac{\varepsilon + \xi - \gamma q_i}{\left( \varepsilon + \xi - \gamma q_i \right)^2 + \gamma^2\left( \frac{1}{2} + q_r \right)^2  } \right], \nonumber\\
	I_i(-\varepsilon) &=& \int_{-\infty}^{\infty} d\xi g(\xi)\frac{-\varepsilon - \xi - \gamma q_i}{\left(-\varepsilon - \xi - \gamma q_i \right)^2 + \gamma^2\left( \frac{1}{2} + q_r \right)^2  },\\
	&=& -\int_{0}^{\infty} d\xi g(\xi)\left[ \frac{\varepsilon + \xi + \gamma q_i}{\left( -\varepsilon - \xi - \gamma q_i \right)^2 + \gamma^2\left( \frac{1}{2} + q_r \right)^2  } +  \frac{-\varepsilon + \xi - \gamma q_i}{\left( \varepsilon - \xi + \gamma q_i \right)^2 + \gamma^2\left( \frac{1}{2} + q_r \right)^2  } \right]. \nonumber
\end{eqnarray}
\end{widetext}
Its obvious that , $ I_i(\varepsilon) = -I_i(-\varepsilon) $ if $ q_i = 0 $, which implies that $ q_i \equiv 0 $. Then, the saddle-point equation becomes
\begin{eqnarray}
	q_r &=& \int_{-\infty}^{\infty} d\xi g(\xi) \int\frac{d\varepsilon}{2\pi} \frac{\gamma \left( \frac{1}{2} - q_i \right)}{\left( \varepsilon - \xi \right)^2 + \gamma^2\left( \frac{1}{2} + q_r \right)^2  },\nonumber\\
	&=& \int_{-\infty}^{\infty} d\xi g(\xi) \frac{\pi}{2\pi}= \frac{1}{2}.
\end{eqnarray}
As the result, we have $ q^R = \frac{1}{2} $. Similarly, its easy to find that $ q^A = -\frac{1}{2} $.

Finally, the saddle-point solution of fermionics gases under dephasing is
\begin{equation}
	Q_0 = \frac{1}{2}
	\begin{pmatrix}
		1 & 2F_\varepsilon \\
		0& -1
	\end{pmatrix},
\end{equation}
where $ 2F_\varepsilon $ is determined by the boundary conditions. The saddle-point of dephasing system is different form that of disorder, it is frequency-independent, that it would results in a finite lifetime to the Fermions and leaves the energy-level invariant.
Because the dynamics are determined by the diagonal part of $ Q_0 $, then we denote the saddle-point as $ Q_0 = \frac{\tau_3}{2} $ for simplicity.
And the corresponding dressed single-particle Green's function is
\begin{equation}
	G^{R/A} = \frac{1}{\varepsilon-\xi_{\mathbf{k}}\pm i\gamma}.
\end{equation}
Then, one can expects that the diffusive nature of density fluctuations in the impurity medium exists in the dephasing system. Notablely, the dephasing is a inelastic scattering process, then the cooperon channel is forbidden due to the absence of time-reversal symmetry. Therefore, the weak/anti-localization phenomena is not expected during the dephasing process.
Nevertheless, general diffusive structure is identical to the disorder system. 

\section{The $ \theta $-term of the dephasing Qi-Wu-Zhang model}
It is known that the effective theory of quantum Hall system is the Pruisken's action, or NL$ \sigma $M with $\theta$-term, which is an effective action of Goldstone modes that emerged when the symmetry of system is broken spontaneously. This technique has been employed in the disordered QWZ model \cite{disorder_qwz}, we will now extend it to account for dephasing QWZ model.
In the dephasing fermionic gases, it is obvious that the matrix mean-field  $ Q^{\alpha\beta}(t,t) = \{\Psi_\alpha(t)\bar{\Psi}_\beta(t)\}$ is unitary symmetric, i.e., $ UQU^\dagger $,  where $ U\in U(2) $.
However, the saddle-point solution $ Q_0 $ is invariant only if $ U\in U(1)\times U(1) $, then the target manifold of soft modes (Goldstone modes) which perturb the saddle-points can be represented as a coset $ Q(x) = T(x) Q_0 T(x)^{-1} $, and $ Q^2(x)=\mathds{1} $, substituting it into the effective action of $ Q(x) $ leads to
\begin{eqnarray}
	S[Q(x)] &=& -\text{Tr} \ln\left[ \tilde{G}_0^{-1}(\varepsilon,\mathbf{k}) + i \gamma T(x) Q_0 T^{-1}(x) \right],\\
	&=&-\text{Tr} \ln\left\{G^{-1}(\varepsilon,\mathbf{k}) + \left[ T^{-1}(x), \tilde{G}_0^{-1}(\varepsilon,k) \right] T(x)\right\}.\nonumber
\end{eqnarray}
where $ G^{-1}(\varepsilon,\mathbf{k}) $ is the dressed single-particle Green's function that
\begin{equation}
	G(\varepsilon,\mathbf{k}) = \begin{pmatrix}
		G^R(\varepsilon,\mathbf{k})&0\\
		0&G^A(\varepsilon,\mathbf{k})
	\end{pmatrix} = \begin{pmatrix}
		\frac{1}{\varepsilon - \xi_{\mathbf{k}}+i\gamma} & 0\\
		0 & \frac{1}{\varepsilon - \xi_{\mathbf{k}}-i\gamma}
	\end{pmatrix},
\end{equation}
where the Keldysh Green's function $ G^{K}(\varepsilon,\mathbf{k})  = F_\varepsilon \circ G^{R}(\varepsilon,\mathbf{k}) - G^{A}(\varepsilon,\mathbf{k})\circ F_\varepsilon $ is determined by the boundary condition $ F_\varepsilon $, and it contributes no dynamical effect in our discussions.

Then, we can approximate the commutator $ \left[ T^{-1}(x), \tilde{G}_0^{-1}(\varepsilon,k) \right] $ with Wigner-Moyal expansion, that $ [X,Y] \simeq -i\{X,Y\} $, where $ \{\cdot,\cdot\} $ is the Poisson bracket.
For QWZ model, $ H(\mathbf{k}) = \vec{h}(\mathbf{k}) \cdot \vec{\sigma} =h_a \sigma^a $, its easy to find that
\begin{equation}
	\left[ T^{-1}(x), \tilde{G}_0^{-1}(\varepsilon,k) \right] T(x) \simeq F_i \Phi_i - \frac{1}{2} J_{ij} \Phi_i \Phi_j,
\end{equation}
with $ F_i = i\partial_{k_i} h_a \sigma^a $, $ J_{ij} = \partial_{k_i} \partial_{k_j} h_a\sigma^a $, $ \Phi_i = \left( \partial_{x_i} T^{-1}\right) T $.
Then, substituting it into the effective action of $ Q(x) $ and expand $ \ln(\cdots) $ up to second-order leads to 
\begin{eqnarray}
	S[Q] &=&-\text{Tr} \ln\left\{G^{-1}(\varepsilon,k) + F_i\Phi_i - \frac{1}{2}J_{ij} \Phi_i\Phi_j \right\}\\
	&=& -\text{Tr} \ln G^{-1}  - \text{Tr}\ln \left\{1 + GF_i\Phi_i - \frac{1}{2}GJ_{ij} \Phi_i\Phi_j \right\}\nonumber\\
	&\simeq&  - \text{Tr}\left( GF_i\Phi_i - \frac{1}{2}GJ_{ij} \Phi_i\Phi_j \right)  + \frac{1}{2} \text{Tr} \left( GF_i\Phi_i \right)^2\nonumber
\end{eqnarray}
where the constant term $ \text{Tr} \ln G^{-1} $ can be omitted.
The algebra is similar as Ref. \cite{disorder_qwz}, we find that the $ \theta $-term in the effective action is
\begin{eqnarray}
	S_{\theta} &=& S_{\theta}^{(1)} + S_{\theta}^{(2)},\\
	S_{\theta}^{(1)}&=& -2i \int_{E}^{\infty} d\varepsilon \int dx \frac{dk_x dk_y}{4\pi^2} \epsilon_{ij} \text{Tr} \left( D_{\varepsilon}^2 \partial_j \Phi_i \right) F_k \nonumber\\
	&=& \frac{\theta_2}{4\pi} \int dx \epsilon_{ij} \text{Tr}\left( \tau_3\partial_i \Phi_j \right)\\
	S_{\theta}^{(2)}&=& -2\gamma \epsilon_{ij} \int dx \frac{dk_x dk_y}{4\pi^2} \text{Tr}\left( D\tau_3 \Phi_i D \Phi_j \right) F_k\nonumber\\
	&=& -\frac{\theta_1}{4\pi} \int dx \sum_s \epsilon_{ij} \text{Tr}\left(P^s\Phi_i P^{\bar{s}}\Phi_j \right) 
\end{eqnarray}
where we have rewritten the Green's function as
\begin{eqnarray}
	G &=& \begin{pmatrix}
		\frac{1}{\varepsilon - h_a \sigma^a +i\gamma} &\\
		& \frac{1}{\varepsilon - h_a \sigma^a -i\gamma}
	\end{pmatrix}\\
	&=& \begin{pmatrix}
		D^+\left[ (\varepsilon + i\gamma) + h_a \sigma^a \right] &\\
		& D^-\left[ (\varepsilon - i\gamma) + h_a \sigma^a \right]
	\end{pmatrix},\nonumber\\
	D^{\pm} &=& \frac{1}{(\varepsilon \pm i\gamma)^2 + h_a h^a},
\end{eqnarray}
and the projector is $ P^{\pm} = \frac{1}{2}\left(\mathds{1} \pm \tau_3 \right) $, that $ D_\varepsilon = D_{\varepsilon}^+ P^+ +  D_{\varepsilon}^- P^-$.
The topological angles $ \theta_1 $ and $ \theta_2 $ is defined as
\begin{eqnarray}\label{theta_angle1}
	\theta_1 &=& 8\gamma   \int \frac{d^2k}{4\pi^2} D_E^+ D_E^- F_k,\\
	\label{theta_angle2}
	\theta_2 &=& 4\pi i \int_{-\infty}^{E} d\varepsilon \int \frac{d^2k}{4\pi^2}\left( D_\varepsilon^{+2} - D_\varepsilon^{-2}\right) F_k,
\end{eqnarray}

\begin{widetext}
	
Furthermore, it is evident that
\begin{eqnarray}
	\mathcal{L}_{\theta}[Q] &=& \epsilon_{ij} \text{Tr}\left[ Q(x) \partial_iQ(x) \partial_j Q(x)  \right], \nonumber\\
	&=& \epsilon_{ij} \text{Tr}\left[ T\tau_3T^{-1} \partial_i(T\tau_3T^{-1}) \partial_j (T\tau_3T^{-1})  \right],\nonumber\\
	&=& \epsilon_{ij} \text{Tr}\left\{ T\tau_3T^{-1} \left[ (\partial_i T)\tau_3T^{-1} + T\tau_3 (\partial_i T^{-1}) \right]\left[ (\partial_j T)\tau_3T^{-1} + T\tau_3 (\partial_j T^{-1}) \right]  \right\}\nonumber\\
	&=& \epsilon_{ij} \text{Tr} \left[ T^{-1}  (\partial_i T) \tau_3 T^{-1} (\partial_j T) + T \tau_3 T^{-1}(\partial_i T) (\partial_j T^{-1})  + (\partial_i T^{-1}) (\partial_j T) \tau_3 + T (\partial_i T^{-1}) T \tau_3 (\partial_j T^{-1})  \right]\nonumber\\
	&=& \epsilon_{ij} \text{Tr} \left[ -T \tau_3 T^{-1} (\partial_j T)   (\partial_i T^{-1}) + T \tau_3 T^{-1}(\partial_i T) (\partial_j T^{-1})  + T \tau_3 T^{-1} (\partial_i T) (\partial_j T^{-1}) - T \tau_3 T^{-1} T  (\partial_j T) (\partial_i T^{-1})  \right]\nonumber\\
	&=& 2\epsilon_{ij} \text{Tr} \left[ T \tau_3 T^{-1}(\partial_i T) (\partial_j T^{-1})  - T \tau_3 T^{-1} (\partial_j T)   (\partial_i T^{-1})  \right],
\end{eqnarray}
And the integral kernel in $ S_{\theta}^{(1)} $ can be rewritten as
\begin{eqnarray}
	4\epsilon_{ij} \text{Tr} \left[ \tau_3 \partial_{i} \Phi_{j} \right] &=& 2\epsilon_{ij} \text{Tr} \left\{ \tau_3 \partial_{i} \left[ \left( \partial_j T^{-1}\right)  T \right] -  \tau_3 \partial_{j} \left[\left(  \partial_i T^{-1}\right)  T \right] \right\} \nonumber \\
	&=& 2\epsilon_{ij} \text{Tr} \left[ \tau_3 \left( \partial_{i} \partial_j T^{-1}\right) T  -  \tau_3 \left( \partial_j T^{-1}\right) \left( \partial_i T\right)  - \tau_3 \left( \partial_j \partial_i T^{-1}\right)  T + \tau_3 \left(  \partial_i T^{-1}\right)  \left( \partial_j  T\right) \right] \nonumber\\
	&=& 2\epsilon_{ij} \text{Tr} \left[ \tau_3 \left(  \partial_i T^{-1}\right)  \left( \partial_j  T\right) - \tau_3 \left( \partial_j T^{-1}\right) \left( \partial_i T\right) \right],\nonumber\\
	&=& 2\epsilon_{ij} \text{Tr} \left[ T\tau_3 T^{-1}\left(  \partial_i T\right)  \left( \partial_j  T^{-1}\right) - T\tau_3 T^{-1}\left(  \partial_j T\right)  \left( \partial_i  T^{-1}\right) \right],\nonumber\\
	&=& \mathcal{L}_{\theta} [Q].
\end{eqnarray}
Similarly, the integral kernel in $ S_{\theta}^{(2)} $ is
\begin{eqnarray}
	4\epsilon_{ij} \sum_{s} \text{Tr} \left[ s P^s \Phi_{i} P^{\bar{s}} \Phi_{j} \right] &=& \epsilon_{ij} \text{Tr} \left[ (1+\tau_3) \left( \partial_i T^{-1} \right) T (1-\tau_3) \left( \partial_j T^{-1} \right) T-  (1-\tau_3) \left( \partial_i T^{-1} \right) T (1+\tau_3) \left( \partial_j T^{-1} \right) T\right],\nonumber\\
	&=& 2\epsilon_{ij} \text{Tr} \left[ \tau_3 \left( \partial_i T^{-1} \right) T \left( \partial_j T^{-1} \right) T-\left( \partial_i T^{-1} \right) T \tau_3 \left( \partial_j T^{-1} \right) T\right],\nonumber\\
	&=& -2\epsilon_{ij} \text{Tr} \left[ T\tau_3 T^{-1}\left( \partial_i T \right)  \left( \partial_j T^{-1} \right) - T \tau_3 T^{-1} \left( \partial_j T \right)\left( \partial_i T^{-1} \right)\right],\nonumber\\
	&=& - \mathcal{L}_{\theta}[Q].
\end{eqnarray}

\end{widetext}

In conclude, the $\theta$-term of dephasing QWZ model is
\begin{equation}
	S_{\theta} = \frac{\theta_1+\theta_2}{16\pi} \int dx \epsilon_{ij} \text{Tr}\left[ Q(x) \partial_iQ(x) \partial_j Q(x)  \right].
\end{equation}
Combining with the diffuson action $ S_{0}[Q] $, the Pruisken's action of dephasing QWZ model is
\begin{equation}\label{pruisken_action}
	S_{\text{eff}}[Q] = S_0[Q] + \frac{\theta}{16\pi} \int d^2x \text{Tr}\left( \epsilon_{ij} Q \partial_i Q \partial_j Q \right).
\end{equation}
The topological angle $\theta = \theta_1+\theta_2$ is parameterized by $(E,u,\gamma)$, i.e., $\theta(E,u,\gamma)$.
Our results in calculating $ \theta $ is identical to the disorder QWZ model in Ref. \cite{disorder_qwz}, however, their results valid only in the weak-disorder regime, in which the self-energy can be approximated with a purely imaginary number and is $ \varepsilon $-independent.
For strong disorder, the self-energy should be evaluated through the SCBA equation of disorder system \cite{Keldysh_book}, which is $ \varepsilon $-dependent and would acquire a finite real part, then Eqs. (\ref{theta_angle1}, \ref{theta_angle2}) breaks down.
Intriguingly, the self-energy in dephasing Fermionic gases is $ i\gamma $, which is $ \varepsilon $-independent, then our results is valid for arbitrary rate of dephasing.

\section{Phase diagram of dephasing Qi-Wu-Zhang model}
In the main part, two fix pints of dephasing quantum anomalous Hall system has been identified with the renormalization group (RG) analysis, the Hall plateaus $(\sigma_{xy},\sigma_{xx})=(n,\sigma_0)$ with $\sigma_0>0$ and the quantum Hall critical points $(\sigma_{xy},\sigma_{xx})=(n+\tfrac{1}{2},\sigma_{xx}^*)$. 
For $ n < \sigma_{xy} < n+1/2 $, the RG at large scales forces $ (\sigma_{xy}, \sigma_{xx}) $ flows to $ (n,\sigma_0) $, while forces $ (\sigma_{xy}, \sigma_{xx}) $ flows to $ (n+1,\sigma_0) $ for $ n+1/2 < \sigma_{xy} < n+1 $.
Then, two phases are expected in the dephasing QWZ model, one is $ |\theta|<\pi $, the other is $ |\theta|>\pi $, and the quantum Hall critical points are those $ |\theta|=\pi $. At large scales, the RG flow sends the topologically nontrivial regime ($|\theta|>\pi$) to the fixed point $|\theta|=2\pi$ and the trivial regime ($|\theta|<\pi$) to the fixed point $|\theta|=0$. 

\begin{figure}[hb]
	\includegraphics[width = 0.5\textwidth]{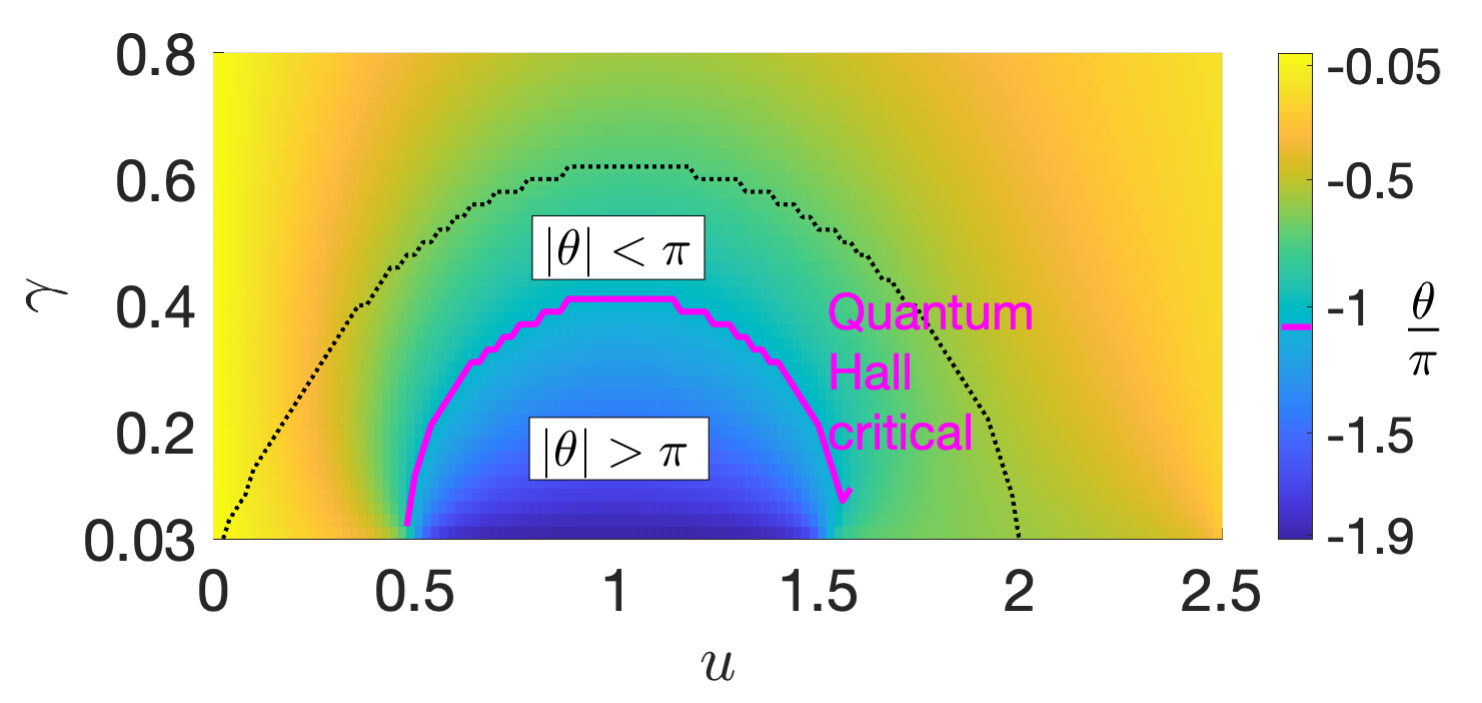}
	\caption{The phase diagram of dephasing Qi-Wu-Zhang model for $ E = 0.5 $, in which there are two phases, denoted as $ |\theta|>\pi $ and $ |\theta|<\pi $ correspondingly, where $ \gamma $ is the rate of dephasing, $ u $ is the staggering potential of the model, $ \theta = \theta_1 + \theta_2 $ is the topological angle. Where the quantum Hall critical points are those $ |\theta|=\pi $ (mengta-solid line), and the black dotted line are critical points for $ E = 0 $.}
	\label{phase_diag_QWZ_0d5}
\end{figure}

The reference energy is usually set at midgap value that $ E=0 $, then $\theta(0,u,\gamma)$ would gives corresponding phase diagram, as presented in Fig. 1 (b) in the main part.
However, the reference energy might deviate from the midgap value in real systems, this will compressing the nontrivial region, see Fig. \ref{phase_diag_QWZ_0d5} of the phase diagram for $ E=0.5 $.

\section{Steady-states of Boundary driven Qi-Wu-Zhang model with dephasing}
xcWe consider the system is driven out of equilibrium by boundary reservoirs and every site subject to dephasing noise, in which the time evolution of the density matrix is the Lindblad master equation \cite{Lindblad_1,Lindblad_2}
\begin{eqnarray}
	\frac{d}{dt} \rho &=& -i[H_{\text{qwz}},\rho] + \mathcal{D}_{1}[\rho] +  \mathcal{D}_{L}[\rho] +  \mathcal{D}_{\text{deph}}[\rho],
\end{eqnarray}
In QWZ model, the sublattice degrees of freedom are denote as $ A $ and $ B $. The dissipators $ \mathcal{D}_{1/L}[\rho] $ describe the action of the two driving baths
\begin{eqnarray}
	\mathcal{D}_{1/L}[\rho] &=&  \Gamma\left(1- f_{1/L}\right) \sum_{i=1/L,j} \left(c_{i,j;A} \,\rho \, c_{i,j;A}^\dagger -\frac{1}{2}\{c_{i,j;A}c_{i,j;A}^\dagger, \rho\} \right), \nonumber\\
	&& +\Gamma f_{1/L}\sum_{i=1/L,j} \left(c_{i,j;A}^\dagger\, \rho\, c_{i,j;A} -\frac{1}{2}\{c_{i,j;A}^\dagger c_{i,j;A}, \rho\} \right) \nonumber
\end{eqnarray}
where $ \Gamma $ is the coupling strength, $ f_{1/L} $ is the Fermi-Dirac distribution of the bath at the left/right side of system that $ \langle c^\dagger c \rangle = f $. For simplicity, we consider the situation which $ f_1 = 1 $ while $ f_L = 0 $, then
\begin{eqnarray}
	\mathcal{D}_{1}[\rho] &=& \Gamma \sum_{i=1,j} \left(c_{i,j;A}^\dagger\, \rho\, c_{i,j;A} -\frac{1}{2}\{c_{i,j;A}^\dagger c_{i,j;A}, \rho\} \right) \nonumber\\
	\mathcal{D}_{L}[\rho] &=& \Gamma \sum_{i=L,j} \left(c_{i,j;A} \,\rho \, c_{i,j;A}^\dagger -\frac{1}{2}\{c_{i,j;A}c_{i,j;A}^\dagger, \rho\} \right).\nonumber
\end{eqnarray}
And $ \mathcal{D}_{\text{deph}}[\rho] $ describes the dephasing on each lattice
\begin{eqnarray}
	\mathcal{D}_{\text{deph}}[\rho] &=& \gamma\sum_{i,j}\left(n_{i,j;A} \,\rho\, n_{i,j;A} -\frac{1}{2}\{n_{i,j;A}, \rho\}\right.\nonumber\\
	&&\left. + n_{i,j;B}\, \rho\, n_{i,j;B} -\frac{1}{2}\{n_{i,j;B}, \rho\} \right).
\end{eqnarray}

Despite the time evolution of density matrix is non-linear due to dephasing, the time evolution of single-particle correlation function $ C_{mn}=\text{Tr}\left( c^\dagger_m c_n \rho \right) $ is linear,  that \cite{steady_corr_func}
\begin{eqnarray}\label{single_evo_t}
	\frac{d}{dt} C &=& -i[\mathbb{H},C] + \gamma\left[ \text{diag}(C) - C \right] \nonumber\\
	&& -\frac{\Gamma}{2}  \left(  M_{L} C + C M_{L} + M_{1} C + C M_{1} - 2 M_{1}  \right),\qquad
\end{eqnarray}
where $ \mathbb{H} $ is the Hamiltonian matrix of the system, $ \left( M_{1} \right) _{m,n} = \delta_{n,i=1 A} $,  $ \left( M_{L}\right) _{m,n} = \delta_{n,i=L_x A} $, that $ \delta_{n,i=1/L_x A} $ is nonzero only if the indices $ n $ equal to the sites of sublattices $ A $ at left/right side ($ i=1/L $).

For the steady-states, $ \dot{C_s}=0 $, then
\begin{eqnarray}
	0 &=& -i[\mathbb{H},C_s] + \gamma\left[ \text{diag}(C_s) - C_s \right] \nonumber\\
	 &&-\frac{\Gamma}{2}  \left(  M_{d} C_s + C_S M_{d} + M_{s} C_s + C_s M_{s} - 2 M_{s} \right),\qquad
\end{eqnarray}
we can rewrite it to a more compact form
\begin{eqnarray}
	-\Gamma M_{1}&=& -i\left(  \mathbb{H}_{\text{eff}} C_s - C_s \mathbb{H}_{\text{eff}}^\dagger \right)  + \gamma \text{diag}(C_s) ,\\
\end{eqnarray}
where $ \mathbb{H}_{\text{eff}} = \mathbb{H} - i \frac{\Gamma}{2}\left( M_{d} + M_{s}\right) - i\frac{\gamma}{2} \mathds{1}.\nonumber $
Vectoring $ C_s $ leads to 
\begin{equation}\label{linear_func}
	\left( -i H_{\text{eff}} \otimes \mathds{1} + i\mathds{1}\otimes H_{\text{eff}}  + \gamma D \right) \cdot \text{vec}\left( C_s \right)  =  -\Gamma \text{vec}\left( M_1 \right),
\end{equation}
where $ D $ is the matrix representation of operator $ \text{diag}(C_s) $.
Then, $ C_s $ becomes a solution of the linear function in Eq. (\ref{linear_func}), i.e., $ \mathbb{L}x=b $, where $ \mathbb{L}= -i H_{\text{eff}} \otimes \mathds{1} + i\mathds{1}\otimes H_{\text{eff}}  + \gamma D $, $ x = \text{vec}\left( C_s \right) $, $ b =  -\Gamma \text{vec}\left( M_1 \right)$, which is a well-defined problem. However, the matrix $ A $ is usually extremely large, for example, $ \mathbb{L} $ is $ 4L^4\times4L^4 $ matrix for QWZ model on square lattice that $ L_x = L_y = L $. The complexity to directly solving Eq. (\ref{linear_func}) is usually $ O(n^3) $-problem, where $ n=4L^4 $ is the scale of matrix $ \mathbb{L} $, that is time-consuming and usually impossible when the scale of system is large.

\begin{table}[htb]
		\centering
		\label{ode45_Krylov}
		\caption{The algorithm of ODE45 (Runge-Kutta method) assisted Krylov-subspace}
		\setlength{\tabcolsep}{3mm}{
			\begin{tabular}{ c }
				\hline
				\hline
				\textbf{1}: Solving the Sylvester function \\
				 \( -\Gamma M_{1}=-i \mathbb{H}_{\text{eff}} \tilde{C}_s +i\tilde{C}_s \mathbb{H}_{\text{eff}}^\dagger  \) .\\
				\textbf{2}: Long time evolution of $ C_t $with ODE45,  where $ C_0 = \tilde{C}_s $, \\
				$ \frac{d}{dt} C = -i\mathbb{H}_{\text{eff}}C + iC\mathbb{H}_{\text{eff}}^{\dagger}  + \gamma\text{diag}(C) + \Gamma M_{1} $, \\
				\textbf{3}: Solving the linear equation in the Krylov subspace \\
				$ \left( -i H_{\text{eff}} \otimes \mathds{1} + i\mathds{1}\otimes H_{\text{eff}}  + \gamma D \right) \cdot \text{vec}\left( C_s \right)  =  -\Gamma \text{vec}\left( M_1 \right) $.\\
				the initial condition of iteration is $ x_0 = \text{vec}(C_t) $,\\
				the \textit{relative residual} is $ 10^{-10} $\\
				\hline
				\hline
			\end{tabular}
		} 
\end{table}

An alternative scheme is by considering the sparse matrix, which simplify the problem to $ O(n) \sim O(n^3) $-problem, then Eq. (\ref{linear_func}) becomes solvable but time-consuming. In order to simplify the problem further, we use ODE45 (Runge-Kutta method) assisted Krylov-subspace method, which is $ O(m\cdot k) $-problem, where $ m $ is the size Krylov-subspace and $ k $ is the numbers of iteration. The complexity of Krylov-subspace method is depends on the spectrum of $ \mathbb{L} $ and the initial condition of iteration $ x_0 = \text{vec}\left( C_0 \right) $.
In our scheme, $ x_0 $ is obtained with sufficient long time evolution of  single-particle correlation function in Eq. (\ref{single_evo_t}) by using the ODE45 method, in which the initial condition of evolution is $ \tilde{C}_s $, that $ \tilde{C}_s $ satisfy the Sylvester function $ -\Gamma M_{s}=-i \mathbb{H}_{\text{eff}} \tilde{C}_s +i\tilde{C}_s \mathbb{H}_{\text{eff}}^\dagger  $. Details about the algorithm in finding $ C_s $ is presented in Table I, where the \textit{relative residual} in our numerical simulation is $10^{-10}$.

\section{Finite-size scaling analysis of the steady-states}
\begin{widetext}
In the boundary-driven Hall setup, the electrochemical potential $\mu(x,y)$ satisfies Laplace’s equation $\nabla^{2}\mu(x,y)=0$ \cite{source_drain_anomalous_Chern}. Then, the boundary conditions that no-flux through the transverse edges $y=1$ and $y=L$ (e.g. $J_y(x,1)=J_y(x,L)=0$) leads to \cite{boundary_eq_Hall}
\begin{equation}\label{chemical_func}
	\frac{\partial}{\partial y} \mu(x,y)  = \lambda \frac{\partial }{\partial x} \mu(x,y),
\end{equation}
where $ \lambda = \frac{\sigma_{xy}}{\sigma_{xx}} $, with solution \cite{boundary_eq_Hall}
\begin{eqnarray}
	\mu(x,y) &=& \frac{\Delta \mu}{L}\left( \lambda y + x \right)+\sum_{m=1,3,\dots} T_m \left[  \cos \left(\frac{m\pi}{L} y\right)\cosh\left(\frac{m\pi}{L} x\right)  + \lambda \sin \left(\frac{m\pi}{L} y\right)\sinh\left(\frac{m\pi}{L} x\right) \right]\nonumber\\
	&& + \sum_{n=2,4,\dots} U_n \left[  \cos \left(\frac{n\pi}{L} y\right)\sinh\left(\frac{n\pi}{L} x\right)  + \lambda \sin \left(\frac{n\pi}{L} y\right)\cosh\left(\frac{n\pi}{L} x\right) \right].
\end{eqnarray}
\end{widetext}
In Fig. \ref{lambda_u}, finite-size scaling of $\lambda$ for $ \gamma = 0.1, 0.15, 0.2, 0.4 $ are studied, where the scaling parameter $ \beta_\lambda $ is obtained by fitting the numerical data of $\lambda (L)$ with $ \lambda(L) = \beta_\lambda \ln L +\alpha $.
Its obvious that as the increase of $ u $, the sign of $ \beta_\lambda $ change twice for $ \gamma = 0.1$ and $ 0.15 $. the critical values of $ u $ are denoted as $ u_1 $ and $ u_2 $ ($u_1<u_2$), that $ \beta_\lambda>0 $ when $u_1<u<u_2$,  while $ \beta_\lambda<0 $ otherwise. And $ \beta_\lambda<0 $ for all $ u $ when $ \gamma = 0.1$ and $ 0.15 $. This gives the inverse $ \beta $-functions presented in Fig. 2 (c) in main part.

where $ k_n = n\pi/L $, $ \Delta \mu $ is the chemical potential difference between the two baths, $ T_m $ and $ U_n $ are constants, $ m/n $ is odd/even integer.
Interestingly, at the center of sample where $ x/y \simeq \frac{L}{2} $, its easy to find that
\begin{eqnarray}
	\cos \left(\frac{m\pi}{L} y\right)\cosh\left(\frac{m\pi}{L} x\right) \approx 0, \\
	\sin \left(\frac{n\pi}{L} y\right)\cosh\left(\frac{n\pi}{L} x\right) \approx 0.
\end{eqnarray}
Then, electrochemical potential $\mu(x,y)$ takes a linear solution at the center of sample
\begin{eqnarray*}
	\mu(x,y) &\simeq& \frac{\Delta \mu}{L}\left[ \lambda  y + x \right] + \sum_{m=1,3,\dots} T_m  \lambda \sin \left(\frac{m\pi}{2} \right)\sinh\left(\frac{m\pi}{2} \right) \\
	&&+ \sum_{n=2,4,\dots} U_n \cos \left(\frac{n\pi}{2} \right)\sinh\left(\frac{n\pi}{2} \right).
\end{eqnarray*}

In the steady-states of boundary-driven QWZ model, the distribution of chemical potential $ \mu(x,y) $ is actually the local densities of particle $  n(x,y) $, $  n(x,y) =  n_A(x,y) + n_B(x,y) $, where $ n_{A/B}(x,y) $ is the diagonal elements of correlation matrix $ C_s $.
And at center regime of the model, we have a linear solution of $\mu(x,y)$, so the coarse-graining value of gradients $\mu_x\equiv \partial_x\mu$ and $\mu_y\equiv \partial_y\mu$ can be evaluated at finite sizes
\begin{eqnarray}
	\mu_x(x,y) &=& \frac{n(x+1, y ) - n(x-1, y )}{2}, \\
	\mu_y(x,y) &=& \frac{n(x, y+1 ) - n(x, y-1 )}{2},
\end{eqnarray}
where $ \frac{L}{4}< x/y \leq \frac{3L}{4} $. Then $ \lambda(x,y) = \mu_y(x,y)/\mu_x(x,y) $, and $ \lambda $ is equal to the average value of $ \lambda(x,y) $
\begin{equation}
	\lambda = \frac{4}{L^2} \sum_{x,y} \lambda(x,y).
\end{equation}

\begin{figure}[htb]
	\includegraphics[width = 0.5\textwidth]{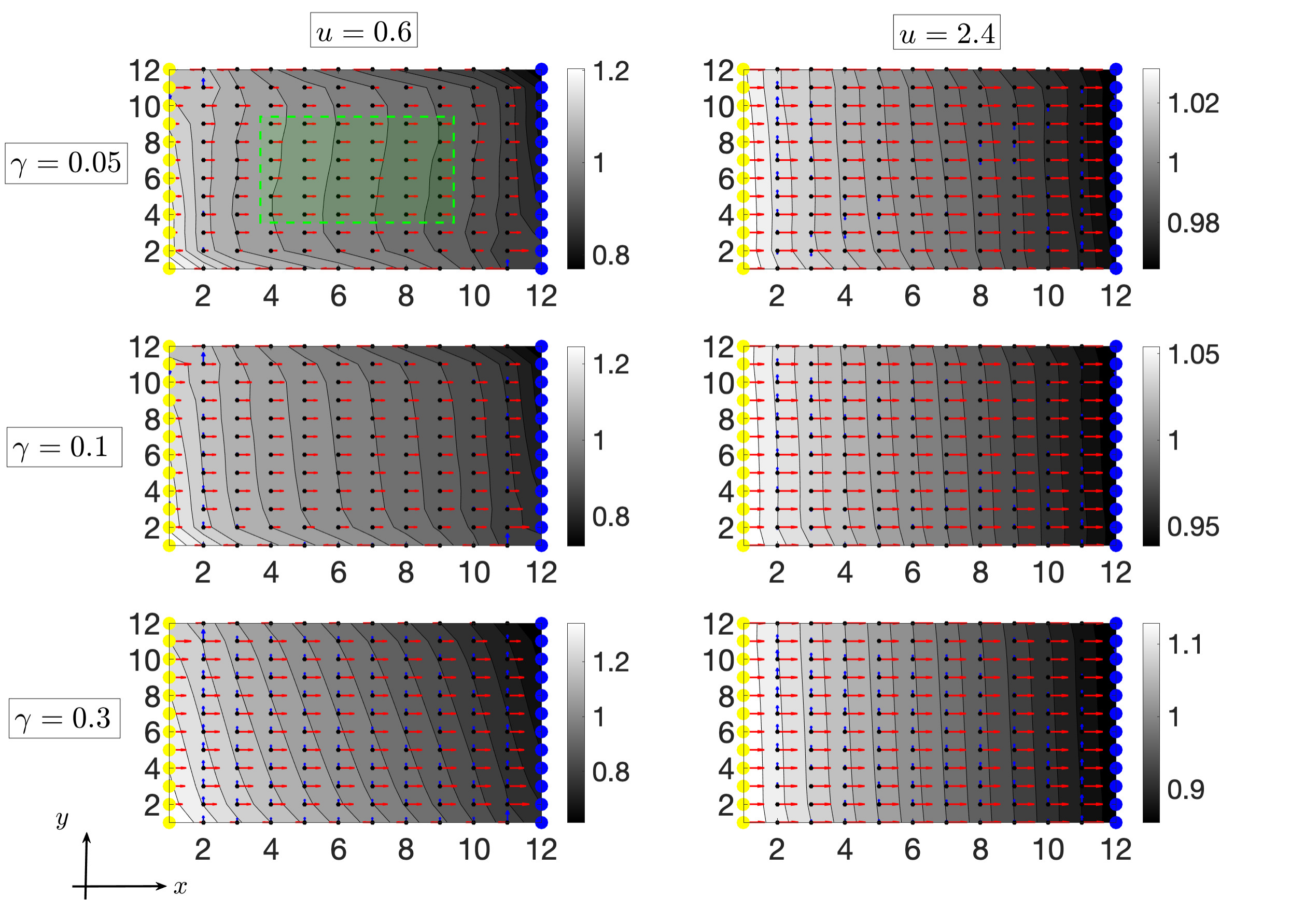}
	\caption{The contour plot of steady-state chemical potential $ \mu(x,y) $ for $ u=0.6 $ (the first column) and $ 2.4 $ (the second column), three values of dephasing rate are studied, $ \gamma=0.05 $ (the first line), $ 0.1 $ (the second line), and $ 0.3 $ (the third line). Where $ \Gamma = 0.1 $, source/drain is marked in yellow/blue, the green region at the center of sample where $ L/4< x/y \leq 3L/4 $ are selected to evaluate $ \mu_x $ and $ \mu_y $. }
	\label{density_distribution}
\end{figure}

\begin{figure*}[htb]
	\includegraphics[width = 1\textwidth]{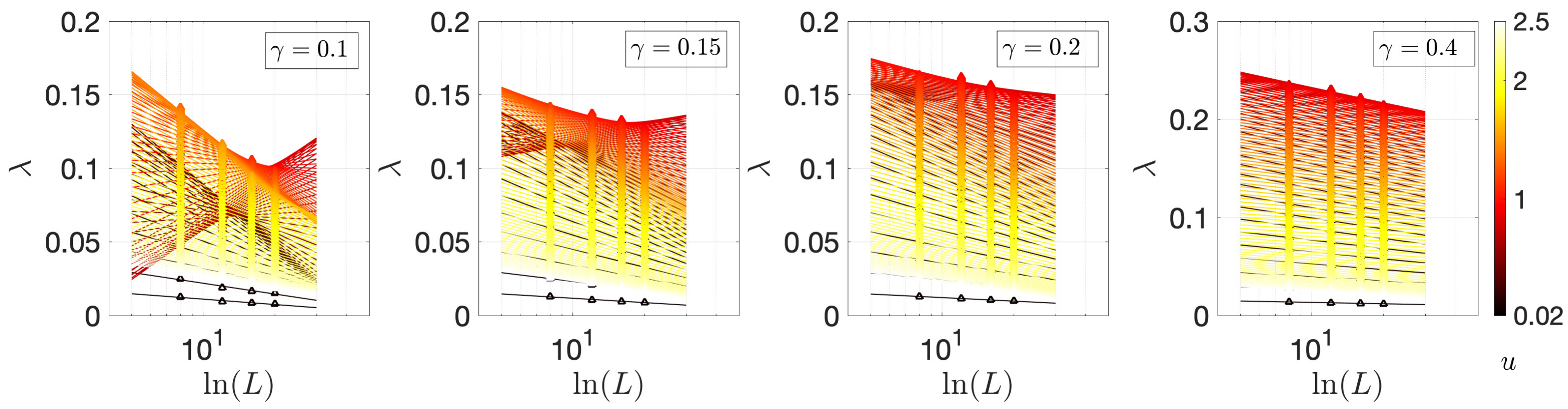}
	\caption{Finite scaling of $ \lambda $ for $ \gamma = 0.1, 0.15, 0.2, 0.4 $, where triangles mark the numerical value of $ \lambda $ for each values of $ u $, and the solid lines are the corresponding fitting curve that $ \lambda(L) = \beta_\lambda \ln L +\alpha $. }
	\label{lambda_u}
\end{figure*}

\begin{figure*}[htb]
 	\includegraphics[width = 1\textwidth]{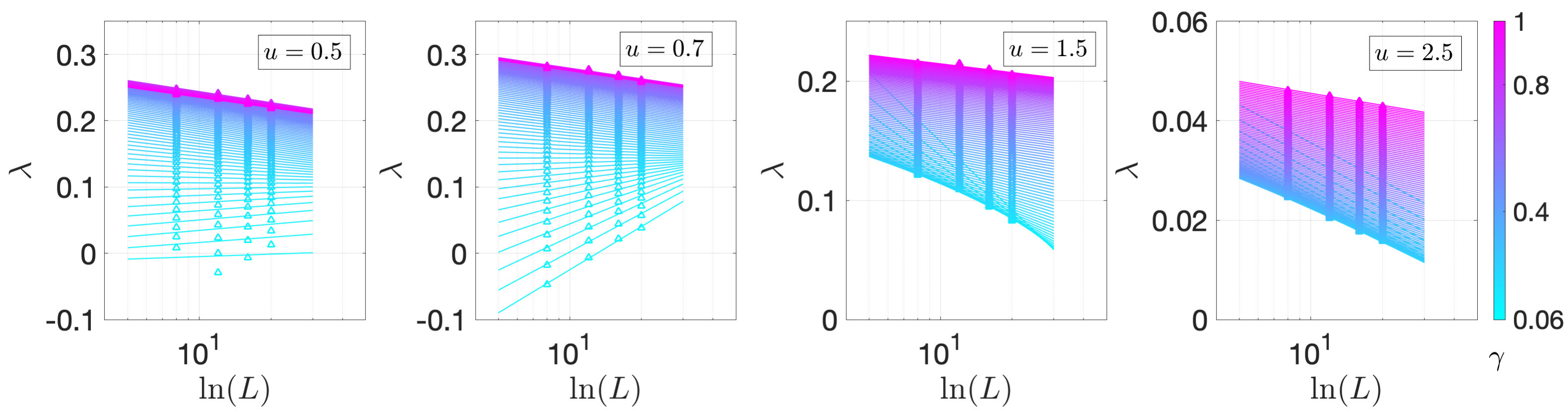}
	\caption{Finite scaling of $ \lambda $ for $ u = 0.5, 0.7, 1.5, 2.5 $, where triangles mark the numerical value of $ \lambda $ for each values of $ \gamma $, and the solid lines are the corresponding fitting curve that $ \lambda(L) = \beta_\lambda \ln L +\alpha $.}
	\label{lambda_gamma}
\end{figure*}

In Fig. \ref{lambda_gamma}, finite-size scaling of $\lambda$ for $ u = 0.5, 0.7, 1.5, 2.5 $ are studied, similarly, the scaling parameter $ \beta_\lambda $ is obtained by fitting the numerical data of $\lambda (L)$ with $ \lambda(L) = \beta_\lambda \ln L +\alpha $.
For $ u = 0.5$ and $ 0.7 $, The sign of $ \beta_\lambda $ changes as the increase of $ \gamma $. The critical values of $ \gamma $ is denoted as $ \gamma_c $, that $ \beta_\lambda>0 $ when $\gamma<\gamma_c$,  while $ \beta_\lambda<0 $ otherwise. And $ \beta_\lambda<0 $ for all $ \gamma $ when $ u = 1.5$ and $ 2.5 $. This gives the inverse $ \beta $-functions presented in Fig. 2 (d) in main part.
One thing to note is that, for $ u = 0.5$ and  $0.7$, $ \lambda <0 $ when $ \gamma $ is pretty small, which is unphysical. This can be attribute to the break down of linear solution, in which the coarse-graining value of $ \lambda $ can't be evaluated on finite sizes, see Fig. \ref{density_distribution} for $ u=0.6 $ and $ \gamma=0.05 $.

\end{document}